# Aplicación de software matemático en carreras de ingeniería

*Application of Mathematical Software in Engineering Careers*

*Aplicação de software matemático em carreiras de engenharia*

**Guillermo José Navarro del Toro**
Universidad de Guadalajara, Centro Universitario de los Altos, México
guillermo.ndeltoro@academicos.udg.mx
https://orcid.org/0000-0002-4316-879X

**Resumen**

Con el objetivo de encontrar vías que puedan conducir a solucionar el problema del aprendizaje de las ciencias exactas e involucrar al estudiante universitario de manera participativa y activa durante el periodo semestral con ayuda de las nuevas tecnologías, se desarrolló la siguiente metodología: *a)* se buscaron aplicaciones (*apps*) para *smartphones* que permitan cálculos; *b)* seleccionadas las *apps*, se identificaron las características para el *software*; *c)* se seleccionaron grupos en los que se pudiera probar el nuevo método; *d)* se desarrolló un banco de problemas para que los participantes los resolvieran, y *e)* se aplicó una encuesta a los estudiantes al concluir los cursos. Como resultado de esta búsqueda, se encontró que existen paquetes de *software* que pueden ser utilizados para las materias de física y que, además, cumplen con el requisito de no tener costo (por lo menos en su versión básica). Después de haber aplicado el nuevo método en tres grupos, en dos periodos semestrales diferentes y en cursos distintos, es halagüeño ver que los alumnos se interesaron por realmente aprender las ciencias exactas, e incluso por gestionar su autoaprendizaje. Los participantes fueron capaces de desarrollar una cantidad de problemas superior a la que se suele resolver con el método convencional.

**Palabras clave:** aplicaciones, ciencias exactas, dispositivo inteligente, docentes.

## Abstract

With the aim of finding ways that can lead to solving the problem of learning the exact sciences and involving the university student in a participatory and active way during the semester period with the help of new technologies, the following methodology was developed: *a)* applications (apps) for smartphones that allow calculations were sought; *b)* selected the apps, the characteristics for the software were identified; *c)* groups were selected in which the new method could be tested; *d)* a bank of problems was developed for the participants to solve, and *e)* a survey was applied to the students at the end of the courses. As a result of this search, it was found that there are software packages that can be used for physics subjects and that, in addition, meet the requirement of being free (at least in their basic version). After having applied the new method in three groups, in two different semester periods and in different courses, it is gratifying to see that the students were interested in really learning the exact sciences, and even in managing their self-study. The participants were able to develop a higher number of problems than is usually solved with the conventional method.

**Keywords:** applications, hard sciences, smart devices, teachers.

## Resumo

A fim de encontrar caminhos que possam levar à solução do problema de aprendizagem das ciências exatas e envolver o universitário de forma participativa e ativa durante o semestre com o auxílio das novas tecnologias, foi desenvolvida a seguinte metodologia: a) buscou-se aplicativos (apps) para smartphones que permitem cálculos; b) selecionados os aplicativos, foram identificadas as características do software; c) foram selecionados grupos nos quais o novo método poderia ser testado; d) foi desenvolvido um banco de problemas para os participantes resolverem; ee) foi aplicada uma pesquisa aos alunos ao final dos cursos. Como resultado dessa pesquisa, constatou-se que existem pacotes de software que podem ser utilizados para disciplinas de física e que, além disso, atendem ao requisito de serem gratuitos (pelo menos em sua versão básica). Depois de ter aplicado o novo método em três turmas, em dois períodos semestrais distintos e em diferentes cursos, é gratificante constatar que os

## Introducción

Todas las herramientas educativas que en algún momento han permitido hacer mejoras en las instituciones de educación superior han evolucionado o han sido sustituidas por nuevas tecnologías aún más eficaces. Actualmente, a raíz del vertiginoso desarrollo tecnológico de los últimos años, nos encontramos frente al reto de actualizar las prácticas y los métodos de enseñanza-aprendizaje que han sido empleados durante mucho tiempo.

El ámbito educativo no solo se ha visto beneficiado por los desarrollos tecnológicos en esta era digital, sino que ha habido varios inventos a lo largo de la historia que han facilitado el aprendizaje dentro de las aulas. Cajori (1909), por ejemplo, documenta el uso de la regla de cálculo en escuelas, institutos y facultades de ingeniería a través del tiempo. Efectivamente, dicha herramienta ha ido cambiando a lo largo de las décadas, y con ella los propios estudiantes. Así, por ejemplo, hasta antes de la década de 1970, era común identificar a un estudiante de ingeniería por portar la regla de cálculo; posteriormente, con la aparición de las calculadoras electrónicas, a partir de la década de 1980, se empezaron a distinguir dos tipos de estudiantes de ingeniería: los que tenían para adquirir una calculadora electrónica (que normalmente eran estudiantes de clases media-alta y alta) y los que seguían empleando la regla de cálculo (generalmente de clases media-baja y baja).

Las calculadoras se fueron haciendo más sofisticadas al ir incluyendo cada vez más funciones, pero sobre todo lo hicieron cuando incluyeron pantallas (*displays*) a través de las cuales se podían ver varias líneas simultáneamente y hasta graficar los resultados. Por tal motivo, los estudiantes con mayor poder adquisitivo pudieron aventajar por mucho a los que no podían adquirir estos dispositivos para resolver problemas de las ciencias exactas en los niveles superior y posgrado.

Afortunadamente, las calculadoras electrónicas conocidas como *calculadoras científicas* entraron en una etapa de producción masiva y las de tipo básico (que en realidad son bastante completas) llegaron a ser accesibles a cualquier bolsillo.

El problema de la enseñanza de las ciencias exactas se fue agudizando por el hecho de que en la clase había una voz que decía lo que se tenía que ir haciendo para resolver los problemas que competían a la sesión correspondiente. Sin embargo, los estudiantes no avanzaban como se deseaba; la práctica docente se reducía a "enseñar" utilizando gis y pizarrón. Se dejó de lado, posiblemente por falta de tiempo, enseñarle al estudiante a usar esas matemáticas fuera del aula o explicarle para qué le servirían prácticamente.

De acuerdo con Muente (2019), en los últimos años, hay grupos de docentes que han buscado, a través de reuniones formales e informales, establecer metodologías que hagan atractivas las ciencias exactas a los estudiantes. Han usado computadoras que, mediante algún paquete de *software*, les ayudan a "explicar" de mejor forma los contenidos de las materias. Desgraciadamente, el estudiante que no tiene una computadora a la mano solo puede tomar nota del método empleado por el docente y, con mucha suerte, podrá reproducir lo que le han mostrado en clase en otro momento.

Otro hecho muy importante es que, en la actualidad, los estudiantes siempre están conectados a sus redes sociales mediante su *smartphone*. De hecho, mucho de su tiempo lo invierten en este tipo de plataformas sociales (aproximadamente 15 % de su tiempo, según Kemp [2020]). Sin duda, se trata de un medio que puede ser utilizado para atraer el interés del estudiante para aprender, comprender y aplicar las ciencias exactas como nunca lo hicieron antes. Y con ello incrementar los índices de aprendizaje y disminuir los de deserción por cambio de carrera.

## Marco teórico

Desde el siglo XVI, científicos han utilizado dispositivos con escalas. La figura más emblemática es Galileo Galilei, quien recurrió a ellos para hacer el cálculo de fórmulas trigonométricas.

Roldán y Sampayo (2015) describen la historia de los logaritmos antes y poco después del descubrimiento del continente americano y apuntan que su inventor fue el matemático Edmund Wingate, a mediados del siglo XVI, mientras que otros se la asignan a William Oughtred, en 1636. Por su parte, Ruiza, Fernández y Tamaro (2004) mencionan que aún tendrían que llegar conceptos como los *números logarítmicos*, de John Napier, cuyos avances fueron incluidos en la regla de cálculo para los cálculos trigonométricos. Sin pasar por alto que el mismo Newton la utilizó para resolver ecuaciones cúbicas, por lo que sugirió se le incorporara un cursor para facilitar las lecturas. Así pues, las distintas contribuciones y sugerencias de varios científicos que trabajaron cálculos diversos moldearon la regla de cálculo. Al respecto, Calvert (19 de enero de 2001) aclara que algunos científicos y matemáticos desarrollaron escalas o sistemas de estandarización para ciertos tipos de cálculos, por lo que hubo constructores de reglas de cálculo que llegaron a contar con sus propias patentes, entre ellos se encuentran Post, Keuffel y Esser y Dietzgen en los Estados Unidos y en Europa la de Dennert y Pape (constructores de la regla Aristo) y Faber, cuyo apellido se posicionó gracias a la marca Faber-Castell.

Hasta inicios de 1980, era muy fácil reconocer a un estudiante de ingeniería porque normalmente iba acompañado de su regla de cálculo. Pero en 1972, de acuerdo con Russo (20 de febrero de 2020), dos egresados de la Universidad de Stanford inventaron la primera calculadora de bolsillo, que se identificó con el nombre de *HP-35* (ver figura 1). Inicialmente, la HP-35 tenía un costo de 390 dólares estadounidenses, un costo poco accesible al bolsillo del estudiante promedio, por lo que solo era accesible a los estudiantes más adinerados, además de que consumía baterías. Fuera de este inconveniente, el avance tecnológico hizo posible que hubiera un auxiliar, de gran facilidad de uso, para realizar cálculos trigonométricos, exponenciales y logaritmos. Su popularidad se incrementó a la par de que sus costos cayeron rápidamente y pronto costaban casi lo mismo que algunas reglas de cálculo, gracias a lo cual se hicieron accesibles a una gran cantidad de gente. En la actualidad, es posible comprar una calculadora científica por menos de 20 pesos y las reglas de cálculo no acompañan más al estudiante de ingeniería.

**Figura 1**. Imagen de la primer calculadora electrónica Modelo HP-35

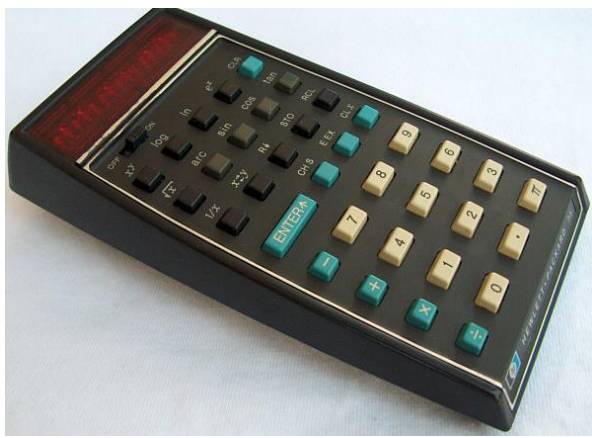

Fuente: HP (s. f.)

Por otro lado, en 1972, se otorga la patente para el primer celular sin alambres a los Laboratorios Bell, quienes, además, tienen la patente del primer módem, siguiendo aquí a Britos (2010). Sin embargo, el primer prototipo fue presentado por Motorola en 1973, y hasta 1984 se inicia la comercialización del modelo DynaTAC 8000X, que era un "ladrillo" de un kilo, de grandes dimensiones y su pila le duraba tan solo una hora. Se trata del miembro que inauguró la denominada *primera generación de celulares*. Uriarte (2020) describe que en 1990 llegó la segunda generación. Estos celulares abrieron las puertas a los futuros sistemas de comunicación digital. De ahí llegó la generación 2.5, que se caracterizó por permitir mensajes de texto y mensajería de multimedios.

Posteriormente, a principios de este siglo, llegó la tercera generación, caracterizada por una gran diversidad de adelantos y marcas. El *boom* del celular llegó a todos los países con esta generación, que incluyó los primeros *smartphones*. En este punto la comunicación se incrementó a velocidades nunca antes alcanzadas, así como el desarrollo de aplicaciones muy diversas para distintos tipos de usuarios y sus necesidades.

Otro factor determinante en los avances tecnológicos de impacto mundial fue el descubrimiento de la súper carretera de la información, que dio inicio en 1958. Castromil (26 de marzo 2016) atribuye este adelanto a la compañía telefónica AT&T, la cual creó el Bell 101 *dataset*, el primer módem comercial capaz de transmitir datos digitales sobre una línea telefónica convencional. De ahí en adelante, manifiesta Uriarte (2020), hubo participación de muchas instituciones que poco a poco fueron desarrollando distintas aplicaciones tecnológicas: se llegó a tener la comunicación entre equipos de cómputo, llegaron los virus, la confirmación de la súper carretera de la información (World Wide Web) y el número de equipos conectados a la Red creció de manera increíble.

Con el tiempo, todas las tecnologías se unieron y llegaron a conformar un sistema al que acceden los usuarios a través de un dispositivo móvil (*smartphone*) y conexión a internet. Ahora hay, aproximadamente, 3000 millones de dispositivos móviles, y se prevé que esta cifra siga creciendo, tal y como lo visualiza Fernández (16 de septiembre de 2020). En la figura 2 se presenta el número de usuarios proyectados para el 2021.

**Figura 2**. Cantidad de *smartphones* en el mundo

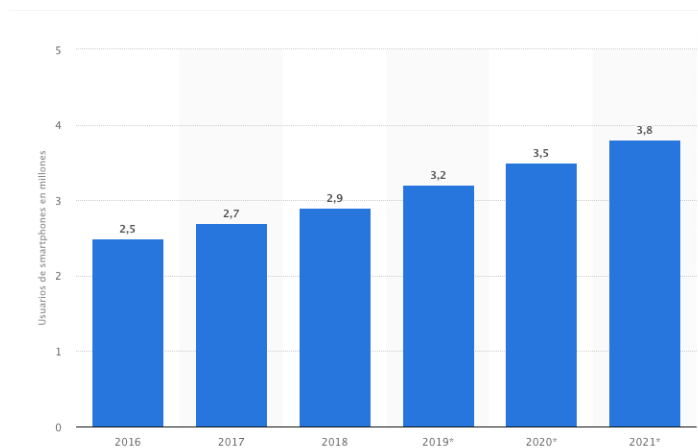

Fuente: Statista (2021)

Otro factor determinante que sigue impactando en el uso de las tecnologías de la información y comunicación (TIC) a nivel mundial ha sido el desarrollo de las aplicaciones, las cuales han crecido de manera increíble en los últimos años, incluyendo las redes sociales, que agrupan las mayores cantidades de usuarios para publicar o ver publicaciones, para compartir información. Un caso que debe ser igualmente mencionado por separado de todos los demás desarrollos que se han realizado en el universo de las TIC es el gran número de aplicaciones creadas para facilitar la educación (a distancia o presencial). El celular inteligente no es solamente un medio para acceder a las plataformas sociales, es una herramienta educativa sumamente poderosa para los estudiantes. Facilita el proceso de adquisición de conocimientos que en el futuro serán parte de su profesión.

## Desarrollo

Para encontrar vías que puedan conducir a solucionar el problema del aprendizaje de las ciencias exactas e involucrar al estudiante de manera participativa y activa en las sesiones de la materia durante el periodo semestral, se desarrolló la siguiente metodología.

*a)* Buscar aplicaciones (*apps*) para *smartphones* que permitan cálculos.
*b)* Seleccionadas las *apps*, identificar las características para el *software*.
*c)* Seleccionar o tomar los grupos en los que se probará el nuevo método.
*d)* Hacer un banco de problemas resueltos por los estudiantes del curso.
*e)* Aplicar una encuesta a los estudiantes al concluir los cursos.

Como primera parte de este nuevo proceso, se buscaron *apps* que cumplieran con los siguientes criterios: *1)* que puedan ser empleadas durante el curso y *2)* que sean de uso libre para no ocasionar gastos al estudiante.

Como resultado de esta búsqueda, se encontró que existen paquetes de *software* que pueden ser utilizados para la materia de Física y que, además, cumplen con el requisito de no tener costo (por lo menos en su versión básica).

Entre los hallazgos, se encuentra la *app* Squid, la cual tiene las siguientes funciones:

- Toma apuntes naturalmente con el lápiz y borra con el dedo.
- Convierte los gráficos en vectores.
- Soporta diferentes estilos y tamaños de hoja.
- Hacer/deshacer, seleccionar, mover y cambiar tamaño.
- Cambia color y grosor de objetos seleccionados.
- Corta, copia y pega entre apuntes.
- Organiza apuntes en libretas.
- Importa, corta y cambia el tamaño de las imágenes.
- Exporta en PDF, PNG o JPEG para imprimir, guardar o compartir.
- Permite compartir apuntes por correo electrónico.
- Acceso directo a nuevos apuntes y abrir libretas.
- Paquete de herramientas: marca texto, borrador real de trazos y figuras.
- Servicios de nube: respalda/restaura y exporta apuntes en PDF a la nube

(Enrique MP, 20 de octubre de 2016).

Otra *app* que puede ser empleada para solucionar problemas en donde intervienen las ciencias duras es handyCalc (Cummins, 24 de marzo de 2011). Dicha herramienta se sugiere para los estudiantes que tienen dificultades al momento de ingresar la información en calculadoras tradicionales, es decir, aquellos estudiantes que tienen dificultades para visualizar el algoritmo matemático que se debe de emplear para solucionar una operación específica o una ecuación sencilla o compleja.

Una aplicación más que cubre los requisitos necesarios para emplearse en las materias en donde se ha de probar el nuevo método de enseñanza es ChampCalc (MDF-XL Pages, 2018). Es muy similar a las calculadoras científicas tradicionales (físicas), sin embargo, al ser una *app*, el poder que brinda para la solución de problemas va más allá de las calculadoras físicas, lo que, combinada con handyCalc, resulta una herramienta de aprendizaje que permite ir más allá de lo que se podría realizar con una sola.

Respecto a las características que debe de tener el *smartphone*, se consultó al fabricante de cada una de las *apps* y se encontró como recomendaciones que el dispositivo trabajara con el sistema operativo Android y contara con un mínimo de 32 gigabytes.

El nuevo método se trabajó con varios grupos. El grupo inicial fue de 40 estudiantes de 1.er semestre de la carrera de ingeniería en Gestión Empresarial y la materia de Introducción a la Física durante el periodo agosto-diciembre 2018.

El siguiente periodo en que se puso a funcionar el método fue el de agosto-diciembre 2019. Aquí participó un grupo de 3.er semestre con 30 estudiantes de la carrera de Contaduría Pública en la materia de Estadística Básica. En el mismo periodo, se tuvo otro grupo de 35 estudiantes de 5.º semestre de la carrera de ingeniería en Gestión Empresarial y la materia de Estadística Inferencial.

## Proceso para facilitar la comprensión de las matemáticas usando el smartphone y apps

Debido a que en las instituciones educativas de nivel profesional al docente se le estimula emplear herramientas que hagan más interesantes las materias que ha de impartir cada semestre, y debido a que, de manera general, las matemáticas con su solo nombre representan un problema para el estudiante en general, se buscó la forma de hacer menos tediosa la enseñanza de estas y que el estudiante se interesara por ellas al ver que realmente son una forma sencilla no solo de cumplir con los créditos académicos, sino que son la base de muchos de los trabajos que podrán enfrentar en su vida como profesionistas. Así pues, se hizo un primer intento por hacer más interesantes y atractivos los cursos en donde se involucran los conocimientos que siempre han sido de difícil comprensión.

Al dar inicio con este procedimiento, no se contaba con referencias estadísticas de los conocimientos que tienen los alumnos de primer semestre, ya que proceden del nivel medio superior y diferentes instituciones educativas, por lo que se consideró que cada uno de los estudiantes reunía conocimientos distintos y el objetivo era que al final del curso contaran con un conocimiento homogéneo.

Como parte del curso, se les explicó a los estudiantes la forma en que habría de llevarse el curso y se solicitó su anuencia para proceder en esta forma, lo que se consiguió de inmediato.

A continuación, se les preguntó por las cualidades del *smartphone* que utilizan de manera continua. Aquí interesaba saber si tenían las características de *hardware* necesarias para que el *software* pudiese ejecutarse sin problema alguno.

Las respuestas que manifestaron fueron impresionantes: solo dos de ellos tenían el *smartphone* con la capacidad mínima de 32 gigabytes y el resto contaba con memorias de 64 y 128 gigabytes, es decir, que están provistos de equipos mejores que el docente.

En la segunda ocasión (en cada uno de los cursos en los que se puso en práctica el método nuevo), las condiciones fueron las mismas. Tal parece ser que el estudiante en general cuenta con dispositivos de alta gama y así puede hacer con velocidad cualquier consulta en Internet.

A los estudiantes de los grupos de $3.^{er}$ y $5.^o$ semestres se les solicitó ver sus calificaciones previas en las materias en donde aparecen las ciencias exactas. Las calificaciones que obtuvieron en términos generales oscilaban entre 72 y 87, por lo cual se tuvo como meta incrementar este promedio, además de que tomaran gusto por las materias de la carrera.

Cabe aclarar que, debido a que actualmente la instituciones de educación superior cuentan con un enlace a internet que a veces funciona muy bien y otras no, los estudiantes en algunas ocasiones tuvieron que acudir a utilizar los datos de sus *smartphones,* sin que ello repercutiera en la obtención de los resultados de los problemas que se fueron resolviendo como parte del curso.

## El curso

Después de que todos descargaron las *apps* y las instalaron en sus celulares, en la segunda sesión se procedió a hacer la primera prueba del *software* que se iba a utilizar durante el semestre. Utilizando el sistema de videoproyección del aula, se introdujo la primera ecuación a la tableta del docente y los estudiantes fueron viendo la forma en que fue escrita, reprodujeron el evento tal y como lo hizo el docente. El primer ejemplo que se presentó a través del videoproyector fue la función de $2.^o$ grado, como se muestra en la figura 3.

**Figura 3**. Función de 2.º grado con la que se dio inicio el curso

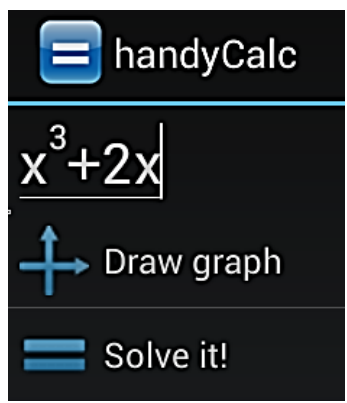

Fuente: Elaboración propia

Los estudiantes inmediatamente se percataron de que era más fácil resolver los problemas empleando el nuevo método, ya que en adelante no tendrían que traer consigo la calculadora científica que la mayoría de ellos había adquirido.

En la figura 4 se muestra el resultado de la función introducida (que es la que se muestra en la figura 3), y con ambas se explicó el significado y se les hizo ver la forma de interpretar los datos, lo que hicieron para comprobar que cualquier función resolverse de forma sencilla y así relacionaron por primera vez una ecuación con los resultados.

**Figura 4**. Gráfica que se desplegó de la función que se introdujo

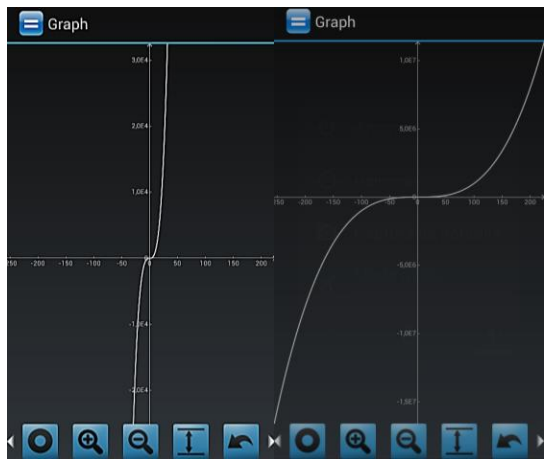

Fuente: Elaboración propia

En la figura 5 se muestra al grupo de estudiantes haciendo el proceso que les mostró el docente. Los alumnos recurrieron a su *smartphone* en las materias que estaban cursando sin utilizar una calculadora científica tradicional.

**Figura 5**. Estudiantes introduciendo la misma ecuación y viendo sus resultados

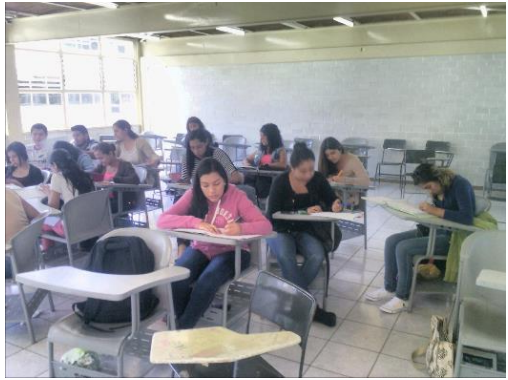

Fuente: Elaboración propia

Como se puede apreciar en la figura 5, los estudiantes se distribuyeron de la forma que les resultaba más apropiada y compartieron los resultados que fueron obteniendo, cosa que no se podía hacer cuando usaban la calculadora científica, de la cual se olvidaron desde el primer día.

**Figura 6**. Estudiantes sustentando una evaluación

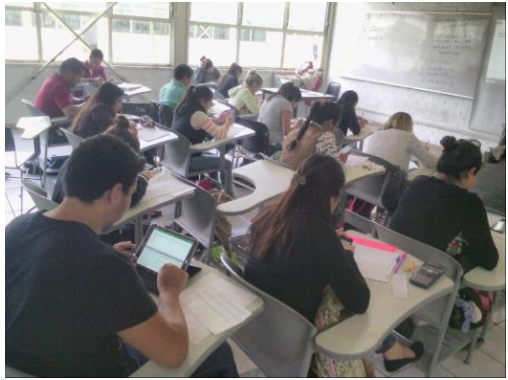

Fuente: Elaboración propia

En la figura 6 se muestra que, al sustentar un examen, el acomodo de las butacas es el que siempre se ha tenido en todas las instituciones educativas, es decir, en filas. En ambas figuras (5 y 6) se puede observar que no hay un solo estudiante que esté utilizando una computadora portátil, solamente su *smartphone*.

También debe mencionarse que los estudiantes comentaron que tenían la costumbre de escuchar al docente cuando iba escribiendo en el pizarrón y explicando lo que hacía y ellos únicamente se dedicaban a copiar y, en algunas ocasiones, hacer algún cálculo en su calculadora, además de copiar lo que se iba escribiendo en el pizarrón.

Con este método se pudieron resolver más problemas de la materia; entendieron la mayoría de los conceptos y sobre todo fueron capaces de relacionar las teorías con sus posibles aplicaciones, es decir, empezaron a aprender ciencias exactas sin emplear una calculadora científica.

Durante el transcurso del semestre, se fue incrementando el nivel de dificultad que tenían los problemas, pues así se especifica en el programa de estudios de la materia.

Por ello, se inició la vinculación de los resultados obtenidos con el *software* instalado en el *smartphone* con las hojas de cálculo, por lo que se les mostró el diseño de los conjuntos de instrucciones que pueden ser ejecutadas de forma secuencial en la hoja de cálculo y que se conocen como *macros*. Como es sabido, una macro tiene la gran ventaja de que puede invocar a otra y esta a otra, por lo que se pueden ir implementando de forma separada y ser almacenadas para ser utilizadas en el problema que así lo requiera. Mediante esta práctica se les pudo explicar cómo diseñar sus propias ecuaciones para que resolvieran los problemas que consideraran los más difíciles dentro de cada tema.

Y con la finalidad de obtener el máximo provecho de las funciones que se pueden desarrollar en Excel mediante el uso de macros, también se utilizó Visual Basic, que permite la creación o mantenimiento de las macros que se hayan creado, pero, sobre todo, que una vez almacenadas puedan invocarse desde cualquier otra macro que vayan instrumentando.

Así, por ejemplo, al estudiante se le proporciona un conjunto de datos como el mostrado en la tabla 1, y se le pide que con su *smartphone*, usando handyCalc y una macro, realice el cálculo de las medidas de tendencia.

**Tabla 1.** Conjunto de datos iniciales

| 32 | 35 | 44 | 40 | 37 | 31 |
| --- | --- | --- | --- | --- | --- |
| 37 | 31 | 35 | 27 | 35 | 29 |
| 35 | 29 | 31 | 32 | 41 | 34 |
| 28 | 25 | 34 | 31 | 31 | 35 |
| 41 | 34 | 29 | 43 | 29 | 31 |

Fuente: Elaboración propia

El resultado que proporciona el estudiante se muestra en la figura 7, en donde se muestran los datos que se van introduciendo y transformando hasta obtener el resultado final.

**Figura 7**. Secuencia de pasos para solucionar el problema propuesto en la tabla 1

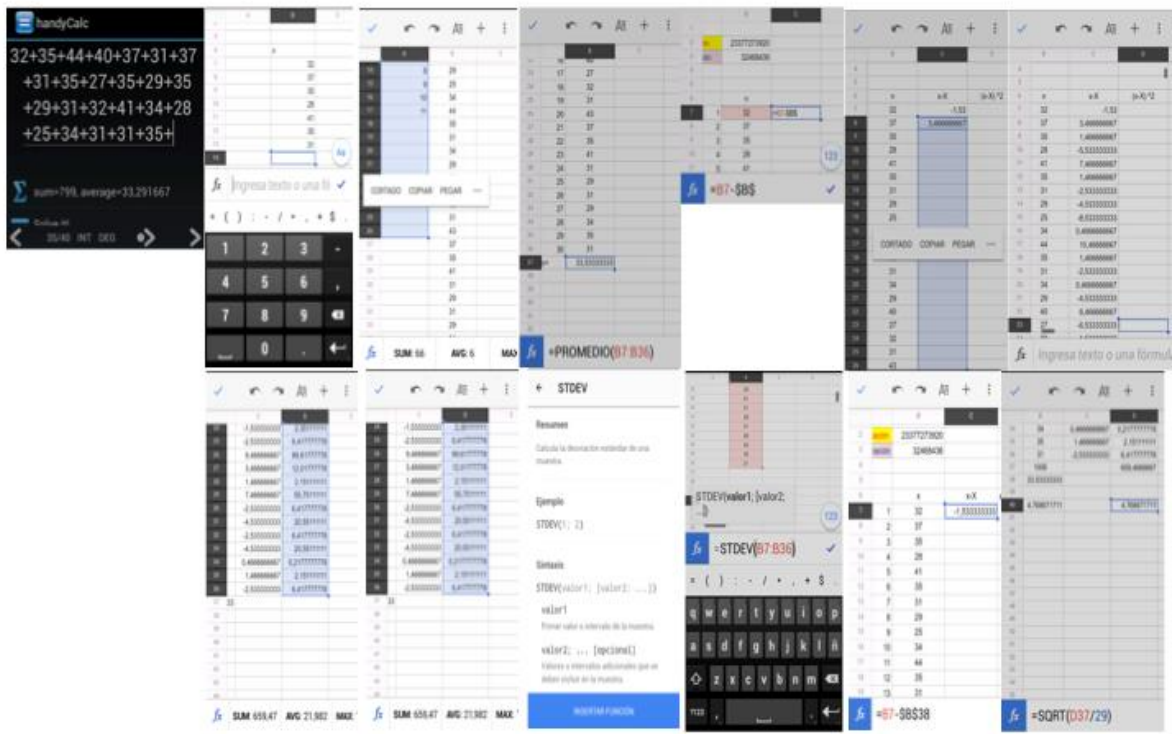

Fuente: Elaboración propia

A la par que se aplicó por segunda vez el método, se estructuró un curso para los docentes del área de las ciencias exactas. Este curso tuvo como principal fin mostrarles que la era de utilizar las nuevas herramientas ha llegado. Se impartió el curso de 40 horas durante el periodo intersemestral, en el mes de enero de 2018. A lo largo de este, se les habló de las bondades que tiene el uso del *smartphone* para enseñar y aplicar las ciencias exactas en el aula de forma directa con los estudiantes y así propiciar que vayan incorporándose a no solo usar el smartphone para entrar a las redes sociales, sino también para aprovecharlo al máximo en sus cursos.

Al inicio, los docentes que estudiaron usando la regla de cálculo y los que usaron la calculadora (siendo mayoría estos, muchos de los cuales fueron aquellos con los que los "nuevos docentes" tomaron sus clases cuando cursaron su carrera de ingeniería en la misma institución, por lo que era una situación muy delicada de abordar, que no sintieran que se trataba de desplazarlos ni mucho menos mostrarles que era hora de que se jubilaran, tampoco para que mostraran una reacción negativa a cambiar lo nuevo por aquello que dio resultados por un periodo de tiempo muy largo) inicialmente creyeron que era un curso más para llenar el espacio de preparación que debe de tener todo docente en sus instituciones educativas.

Por ello, acudieron llevando consigo sus calculadoras, que cuentan con una gran cantidad de funciones y que han empleado durante muchos años. Se procedió a explicar el contenido del curso, se les pidió sacar su *smartphone* y descargar las aplicaciones requeridas (las mismas que se usaron con los estudiantes), y se empezaron las explicaciones tal como sucedió en el primer grupo de estudiantes. Algunos tomaron su calculadora por no creer que se estaban obteniendo los resultados con un número mayor de dígitos de aproximación y empezaron a interesarse por estar actualizados en el uso de estas aplicaciones y con el objetivo de obtener mejores resultados de la enseñanza de las ciencias exactas.

Esto desgraciadamente no lo pudimos constatar, es decir, que lo hayan llevado a la práctica en el periodo agosto-diciembre 2018; y si lo tomaron muy en serio, entonces descubrieron una excelente herramienta para aplicarla en las condiciones actuales, es decir, a distancia.

Al término de los tres cursos que se impartieron utilizando el nuevo método, se les aplicó una encuesta, que se conformó de la siguiente manera:

*1)* ¿Cuál es tu opinión ahora de las matemáticas?
- ◊ No sé
- ◊ Más fácil
- ◊ Son igual

*2)* ¿Cómo son tus resultados con este método?
- ◊ Mejores
- ◊ Peores
- ◊ Iguales

*3)* ¿Recomendarías el método para tus demás cursos?
- ◊ Sí
- ◊ No
- ◊ Dependerá del docente

*4)* ¿Tuviste problemas para comprender el curso?
- ◊ Si
- ◊ No

Se debe remarcar el hecho de que se aplicó siempre al concluir el curso, con la intención de que, ya sabiendo sus calificaciones, pudieran contestar de la forma más honesta posible.

# Resultados

Las encuestas aplicadas a los estudiantes muestran cómo fueron percibidos los cursos entre los tres grupos. Cabe mencionar que ha sido una experiencia profesional muy satisfactoria. El haber iniciado con un trabajo de tipo experimental en favor del aprendizaje está dando pie al inicio de una era (usar aplicaciones a través de *smartphones* en lugar de las costosas calculadoras) y la terminación de otra (que lleguen al desuso costosas calculadoras).

Como parte de los resultados obtenidos por los estudiantes, se debe decir que no hubo reprobados.

Las respuestas a la pregunta uno de la encuesta, que fue "¿Cuál es la opinión que tiene el estudiante sobre las matemáticas después de haber tomado el curso con este método?", se muestran en la figura 8.

**Figura 8**. En la pregunta uno se refleja que la aplicación del método facilitó la comprensión de la materia

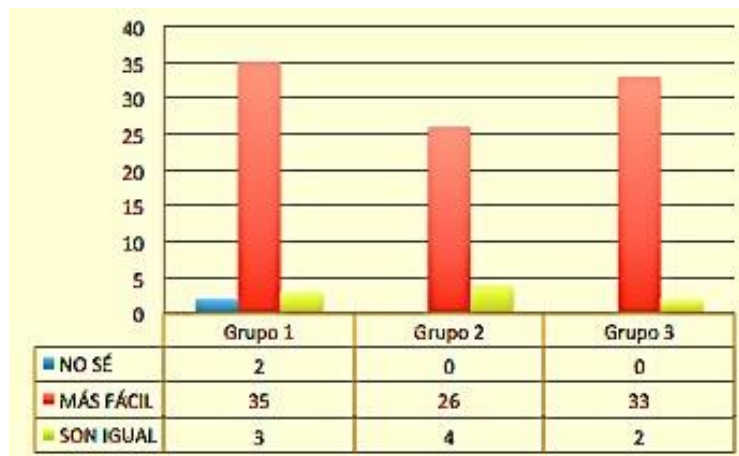

Fuente: Elaboración propia

De ello se concluye que a la gran mayoría de los participantes se les facilitó mucho la forma de llevar el curso y encontraron que no solo pueden usar el *smartphone* para las redes sociales, sino para adquirir los conocimientos que le serán de mucho beneficio y ayuda en su vida profesional. Sin embargo, también se puede observar que en los tres grupos existen estudiantes que siguen considerando que el resultado con cualquier método sería el mismo, y no faltaron los despistados que no saben que puede ser mejor para ellos.

Los resultados de la pregunta dos, que fue "¿Cómo consideran que son los resultados que obtuvieron al emplear este nuevo método de enseñanza?", se observan en la figura 9. Las respuestas obtenidas son a favor de utilizar el método nuevo dejando atrás las costosas calculadoras, tal como lo manifestó la mayoría de los estudiantes.

También hubo respuestas que indican que no hay cambio entre el método tradicional (exposición-explicación-ejercicios-tarea-calculadora) y el aquí usado. Destaca que el primer grupo tuvo a dos estudiantes que afirman que los resultados fueron peores.

**Figura 9**. En la pregunta dos se les cuestionó cómo fueron los resultados que obtuvieron con este método

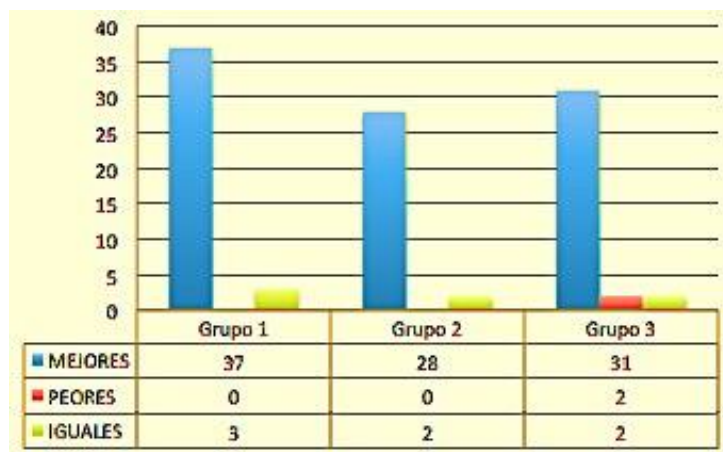

Fuente: Elaboración propia

La pregunta tres, que consistió en saber si los estudiantes estarían dispuestos a que en sus demás cursos en donde aparecen las ciencias exactas se siguiera el mismo método (ver figura 10), arrojó que muchos estudiantes estarían de acuerdo en que los docentes que cubren esas materias se capacitaran para poder impartir cursos usando el nuevo método, lo cual es debido a que resulta más sencillo y atractivo, y destaca el hecho de que alcanzan a cubrir una mayor cantidad de material, con lo que se hace más completo e interactivo su curso.

**Figura 10**. En la pregunta tres se muestra que la mayoría de los estudiantes está dispuesto a continuar con el método en sus demás cursos

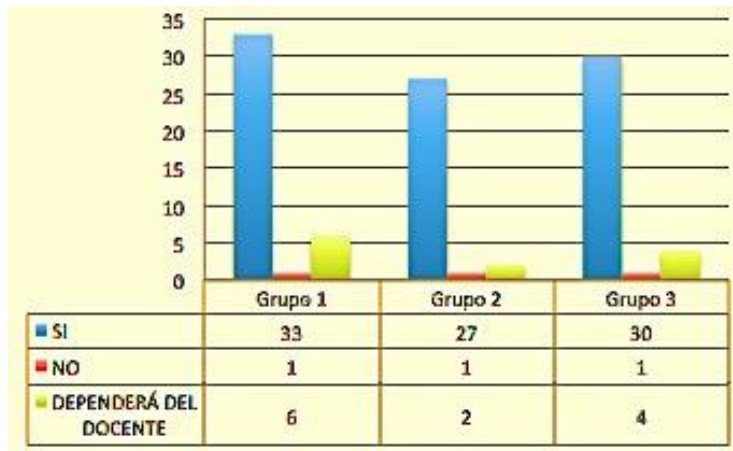

Fuente: Elaboración propia

Desafortunadamente, hubo algunos estudiantes que sin importar que hayan obtenido mayores calificaciones y mejores y más conocimientos se empeñan en regresar al método tradicional.

**Figura 11**. Resultados de la pregunta cuatro sobre la cantidad de conocimientos que adquirió durante el curso

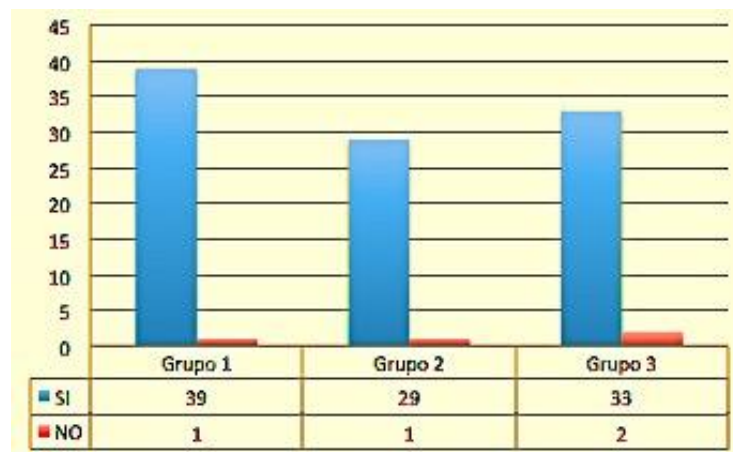

Fuente: Elaboración propia

Por último, en la figura 11 se observan los resultados respecto al cuestionamiento cuatro, que gira en torno a la comprensión del curso. Después de haber aplicado el nuevo método en tres grupos, en dos periodos semestrales diferentes en cursos distintos, es halagüeño ver que los alumnos se han interesado por realmente aprender no solo las ciencias exactas, sino a gestionar el autoaprendizaje. Los participantes fueron capaces de desarrollar una cantidad de problemas superior (250 problemas distintos y no de la clase) a la que se hace con el método antiguo (28 problemas distintos y no de clase) en el de grupo de 35 estudiantes de 5.º semestre de la ingeniería en Gestión Empresarial y la materia de Estadística Inferencial.

También hubo respuestas negativas que nos conducen a hacer mejoras continuas al método, de tal manera que sea un método adaptativo a los cambios en los dispositivos que se empleen, al *software* y a los planes y programas de estudios que forman parte de la retícula.

Ahora bien, por el lado del curso que se impartió a 20 docentes del área de las ciencias exactas, aún desconocemos si han aplicado o no en sus grupos lo aprendido. Esperamos y confiamos que así sea, sobre todo porque en este tiempo en que la educación se está llevando a distancia sería una herramienta que les daría el rendimiento óptimo a ellos y a sus estudiantes.

## Discusión

El haber iniciado una nueva experiencia profesional de manera experimental usando las nuevas tecnologías de la de la información y comunicación puede ser visto como un punto de partida para que el ahora estudiante (profesionista del futuro cercano) vaya perdiendo el miedo a explorar nuevas oportunidades en las que podrá o no poner a prueba sus conocimientos, al igual que ha sucedido con los docentes que dieron inicio con esta nueva forma de impartir las matemáticas.

Tal vez no haya sido la mejor de las posibles formas que pudieran existir para que comprendan las matemáticas, pero sí fue el primer paso (cierto o no) para que el docente se vaya desprendiendo de las formas tradicionales, de ser la única persona que habla durante todo el periodo semestral cuando se está impartiendo un curso de matemáticas.

Tal vez muchos estudiantes habrán despertado y otros no, tal vez muchos otros docentes no estarán de acuerdo con los contenidos y herramientas que se han utilizado en estos cursos, pero todo lo aquí realizado puede ser mejorado, y con la participación de muchos otros docentes y un mayor número de estudiantes se podrán ir haciendo trabajos docentes especializados que permitan beneficiar a un mayor número de participantes.

## Conclusiones

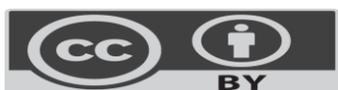

Los estudiantes participantes estuvieron dispuestos a usar el *smartphone* con la intención de aprender más fácilmente y mejorar sus calificaciones. El docente, de manera general, está acostumbrado a impartir sus materias de cierta forma, sin experimentar nuevos métodos, muchas veces replicando el método aplicado cuando él era estudiante. Sin embargo, muchos tuvimos la oportunidad de aprender con docentes que siempre fueron innovadores en su campo y se atrevieron a mejorar las condiciones de aprendizaje de sus estudiantes. En el presente caso, tratamos de no ser estáticos en los métodos de enseñanza y aprendizaje, al igual que algunos docentes con los que tuvimos contacto durante la carrera.

También es necesario mencionar que existen otros docentes que aún se niegan a cambiar su forma de impartir sus enseñanzas y nos pusieron en predicamentos al tratar de involucrarlos en montar nuevos escenarios que favorezcan al estudiante.

## Futuras Líneas de Investigación

El procedimiento que se empleó, la forma en que se desarrolló y los resultados que se obtuvieron se han ido divulgando en distintas instituciones educativas. A muchos de los docentes les resultó muy interesante el poder participar en hacer mejoras y llevarlas a cabo en sus cursos, mientras que otros todavía se muestran incrédulos respecto al procedimiento y a la aplicación de la tecnología en el aula y por lo tanto se abstienen de participar.

Debido a todo lo anterior, se han estado diseñando cursos exclusivos para mostrar a los docentes las bondades de este procedimiento, partiendo de que a algunos de ellos se les tendrá que dar explicaciones desde el uso del *smartphone* que tienen para que puedan participar de manera activa.

Se han estado diseñando cursos usando el *smartphone*, tableta y computadora para ser aplicados en modalidad en línea y a distancia.

En cada uno de los casos, se espera que los participantes nos hagan sugerencias que permitan enriquecer el proceso elegido y así tener una mejora continua.

**Referencias**

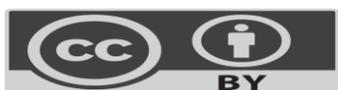